\documentclass[aps,prc,reprint,showpacs]{revtex4-1}
\usepackage{amsmath}
\usepackage{graphicx}
\usepackage{hyperref}
\usepackage{subfigure}

\newcommand{ \be }{\begin{equation}}
\newcommand{ \ee }{\end{equation}}
\newcommand{ \bea }{\begin{eqnarray}}
\newcommand{ \eea }{\end{eqnarray}}
\newcommand{ \la }{\langle}
\newcommand{ \ra }{\rangle}
\newcommand{ \lp }{\left(}
\newcommand{ \rp }{\right)}
\newcommand{ \pt }{$p_t$~}
\newcommand{ \R }{{\cal R}}
\newcommand{ \Qs }{$Q_s^2$}

\begin{document}

\title{Flow Fluctuations from Early-Time Correlations in Nuclear Collisions}

\author{Sean Gavin$^a$ and George Moschelli$^b$}
\affiliation{a) Department of Physics and Astronomy, Wayne State University, 666 W Hancock, Detroit,
MI, 48202, USA\\
b) Frankfurt Institute for Advanced Studies, Johann Wolfgang Goethe University,
Ruth-Moufang-Str.~1, 60438 Frankfurt am Main, Germany}

\date{\today}

\begin{abstract}
We propose that flow fluctuations have the same origin as transverse momentum fluctuations. The common source of these fluctuations is the spatially inhomogeneous initial state that drives hydrodynamic flow. Longitudinal correlations from an early Glasma stage followed by hydrodynamic flow quantitatively account for many features of multiplicity and $p_t$ fluctuation data.  We develop a framework for studying flow and its fluctuations in this picture. We then compute elliptic and triangular flow fluctuations, and study their connections to the ridge. 
\end{abstract}


\maketitle

%
%
\section{\label{sec:intro}Introduction}

Fluctuations in the early stages of nuclear collisions contribute to the anisotropic flow measured in RHIC and LHC experiments, particularly 
the odd harmonics. In Ref.\ \cite{Gavin:2011gr}  we find that this early-time variation can also account for the multiplicity and transverse momentum fluctuations measured in the same experiments. Nevertheless, flow and fluctuation observables reveal different aspects of the initial state.  Harmonic flow is generated largely by the global spatial anisotropy of the initial state. In contrast, fluctuation observables probe local spatial correlations that are largely independent of the overall geometry  \cite{Gavin:2011gr}.  

In this paper we argue that the {\em fluctuations} of the harmonic flow coefficients are driven by the very same local correlations that give rise to $p_t$ and multiplicity fluctuations.  Moreover, if early time correlations are the only source of these fluctuations, then the two measurements are related with no free parameters. 

Experimenters define flow fluctuations by studying the difference of the flow coefficients  $v_n\{2\}$ and  $v_n\{4\}$ measured using two and four particle correlations, respectively \cite{Pruthi:2011eq,Sorensen:2011fb,ALICE:2011ab,BilandzicThesis}.  Jets, resonance decays, and HBT effects also contribute to that difference. Measurements using particles separated in rapidity by more than $1-2$ units eliminate these short range effects. Significantly, a difference between $v_n\{2\}$ and $v_n\{4\}$ remains \cite{Collaboration:2011yba}, suggesting that flow fluctuations are a long range phenomenon. Causality dictates that such long range correlations originate in the early stages of the collision \cite{Dumitru:2008wn,Gavin:2008ev}. 

To compute flow fluctuations we build on an approach started in Refs.\  \cite{Gavin:2008ev,Moschelli:2009tg} in which long range correlations result from the fragmentation of Glasma flux tubes. In this formulation local spatial correlations emerge from fluctuations in the number and distribution of flux tubes. These spatial correlations are then modified by transverse expansion, giving rise to azimuthal correlations. 
A similar physical picture motivates studies using a wide range of different techniques  \cite{Voloshin:2003ud,Pruneau:2007ua,Lindenbaum:2007ui,Sorensen:2008bf,Peitzmann:2009vj,Takahashi:2009na,Andrade:2010xy,Werner:2010aa}. 

This paper is organized as follows. In Sec.\ref{sec:vn}, we begin with a general description of multi-particle correlations and flow coefficients. We relate the intrinsic correlations in the multi-particle system to flow and its fluctuations. To begin, we write expressions for $v_n\{2\}$ and $v_n\{4\}$ in terms of correlation functions, adapting the tools invented in Refs.\ \cite{Borghini:2000sa,Borghini:2001vi} to our framework. Further results are in the appendix. We next define flow fluctuations in terms of $v_n\{2\}$ and $v_n\{4\}$, and turn to discuss the physical interpretation of this definition. 

We discuss the common influence of local correlations on flow and $p_t$ fluctuations in Sec.\ref{sec:loco}. This relationship is the heart of our work. The fluctuations of these quantities are both consequences the spatially inhomogeneous collision environment modified by hydrodynamic flow. The general arguments in Secs.\ \ref{sec:vn} and \ref{sec:loco} set the stage for the more phenomenological  analysis that follows. 

In Sec.\ref{sec:glasma} we describe initial state fluctuations in a CGC-Glasma picture. The number and distribution of Glasma flux tubes relative to the geometrical shape of the system ultimately determine the energy, projectile-mass, and centrality dependence of flow fluctuations. Next, in Sec.\ref{sec:BW} we explain how collective flow and the correlation to the reaction plane modify local correlations. 

We calculate the contribution of long range correlations to the flow coefficients and their fluctuations in Sec.\ref{sec:BW}. Our results are in good agreement with the latest LHC and RHIC data. In Sec.\ref{sec:ridge}, we argue that these same flow fluctuations are also responsible for the ridge. The same factors that influence flow fluctuations also determine the multiplicity and momentum fluctuations computed in \cite{Gavin:2011gr}. We emphasize that no new parameters are introduced here, so that results can be compared directly with \cite{Gavin:2011gr}. 

In Sec.\ref{sec:factotum} we discuss the extent to which the flow coefficients determine the angular distribution of the ridge. We also discuss the possible factorization of the Fourier coefficients of the pair distribution into products of flow coefficients \cite{Adare:2008ae,Aad:2012bu,Chatrchyan:2012wg,Luzum:2011mm}. We show that flow fluctuations can violate factorization at low $p_t$, depending on how the flow coefficients are defined.  We then speculate that factorization holds for momenta $> 1-2$~GeV regardless of the source of anisotropy.  

Lastly, we summarize and discuss the broader implications of our results in Sec.\ref{sec:summary}. 
%
%
%
%
\section{\label{sec:vn}Flow and Its Fluctuations}
Nuclear collisions at non-zero impact parameter $b$ produce anisotropic flow
\cite{Voloshin:2008dg,Sorensen:2009cz}. This anisotropy derives from the change in the shape of the
collision volume with respect to the reaction plane, i.e., the plane spanned by $\mathbf{b}$ and the
beam direction. If the reaction plane is known, this anisotropy is characterized by the moments
\begin{equation}\label{eq:vnRP}
\langle v_n\rangle   =  \langle \cos{n(\phi-\psi_{{}_{RP}})}\rangle,
\end{equation}
where $\phi$ is the azimuthal angle, $\psi_{{}_{RP}}$ is the angle of the reaction plane, and the
brackets denote an average over particles and events. While the reaction plane cannot be observed directly, its
influence can be deduced from multi-particle correlation measurements.  Many strategies have been
employed for measuring these flow coefficients, and they all have strengths and weaknesses. 
 
Experimenters often deduce the flow from the two-particle cumulant
\begin{equation}\label{eq:vn2}
v_n\{2\}^2   =  \langle \cos{n(\phi_1-\phi_2)}\rangle,
\end{equation}
exploiting the reaction plane information implicit in the relative distribution of particle pairs. 
Specifically, they measure the relative azimuthal angle for each particle pair in each event, and
then average over events to obtain  
 \begin{equation}\label{eq:v2FluctExp}
v_n\{2\}^2  = 
\frac{\langle \sum_{i \neq j}\cos{n(\phi_i-\phi_j)}\rangle}
{\langle N(N-1)\rangle}
 \end{equation}
\cite{Bilandzic:2010jr}. 
The cumulant method was developed in Ref.\ \cite{Borghini:2000sa,Borghini:2001vi} and has seen extensive
use \cite{Voloshin:2008dg,Sorensen:2009cz}. 

In this section we cast the framework of  \cite{Borghini:2000sa,Borghini:2001vi} in terms of our
formulation of correlations in \cite{Gavin:2011gr} to study event-wise flow fluctuations. Many
results in this section  are implicit in \cite{Borghini:2000sa,Borghini:2001vi}; only our
application to the computation of fluctuations is new. 
We start by writing (\ref{eq:v2FluctExp}) as
\begin{equation}\label{eq:Dynamic0}
    v_n\{2\}^2 =
    \int\! \,
    \frac{\rho_2(\mathbf{p}_{1},\mathbf{p}_{2})}{\langle N(N-1)\rangle}
    \cos{n(\phi_1-\phi_2)} \, d\mathbf{p}_{1}d\mathbf{p}_{2},
\end{equation}
where the integrals are over the momenta of particles $\mathbf{p}_{i}$ for $i= 1,2$ and the 
distribution of particle pairs is
\begin{equation}\label{eq:pairDensity}
 \rho_2(\mathbf{p}_1,\mathbf{p}_2) = \frac{dN}{d\mathbf{p}_{1}d\mathbf{p}_{2}}. 
\end{equation}
In the absence of correlations, $ \rho_2(\mathbf{p}_1,\mathbf{p}_2) \rightarrow
\rho_1(\mathbf{p}_1)\rho_1(\mathbf{p}_2)$, where the single particle distribution is
$\rho_1(\mathbf{p}) = dN/d\mathbf{p}$.  We stress that the densities $\rho_1$ and $\rho_2$ are
event-averaged quantities that respect the reaction plane. The factorization of $\rho_2$ then
allows for factorization of (\ref{eq:Dynamic0}) such that $ v_n\{2\}^2 \rightarrow \langle v_n\rangle^2$.

Including two-particle correlations, the pair distribution does not factorize.
To identify the contributions of the mean anisotropic flow (\ref{eq:vnRP}) and genuine two-particle
correlations to $v_n\{2\}$, we follow \cite{Borghini:2000sa} and write (\ref{eq:pairDensity}) in
terms of a cumulant expansion,
\begin{equation}\label{eq:corrFunExp0}
    {\rho}_2(\mathbf{p}_1,\mathbf{p}_2) = 
    {\rho}_1(\mathbf{p}_1){\rho}_1(\mathbf{p}_2) +  r(\mathbf{p}_1, \mathbf{p}_2),   
\end{equation}
where $r$ is the two-particle correlation function.  We then use (\ref{eq:vnRP}),
(\ref{eq:Dynamic0}), and (\ref{eq:corrFunExp0}) to write 
\begin{equation}\label{eq:vnsigma}
    v_n\{2\}^2 = \langle v_n\rangle^2 + 2\sigma_n^2. 
\end{equation}
This result is standard, and the factor of two in $\sigma$ is conventional \cite{Voloshin:2007pc}. 
The contribution to (\ref{eq:vnsigma}) from two-body correlations is
\begin{equation}\label{eq:sigmaDef}
    \sigma_n^2 =  
        \int \! d\mathbf{p}_{1}d\mathbf{p}_{2}\,
   \frac{r(\mathbf{p}_{1},\mathbf{p}_{2})}{2\langle N(N-1)\rangle}
    \cos n\Delta\phi,
\end{equation}
where $\Delta\phi = \phi_1-\phi_2$ is the relative azimuthal angle of the correlated particles.

Strictly speaking, this quantity measures the event-wise fluctuations of anisotropic flow due to
geometry and dynamics together with ``non-flow'' correlations from resonance decays, the HBT
effect, and jets. We will not address non-flow effects here. Observe that  all such correlation
effects are of order $1/N$, where $N$ is the number of particles in the system. We focus on
contributions to fluctuations at leading order in $N$. Accordingly, we omit a factor $\langle
N\rangle^2/\langle N(N-1)\rangle \approx 1$ that multiplies $ \langle v_n\rangle^2$ term in
(\ref{eq:vnsigma}); see appendix \ref{appendA} for exact formulae. 

To reduce the effect of fluctuations on the flow signal, experimenters also measure the four particle
cumulant $v_n\{4\}$ defined by the relation  
\begin{eqnarray}\label{eq:v4FluctExp}
v_n\{4\}^4  = 
 2v_n\{2\}^4 -  \langle \cos{n(\phi_1+\phi_2-\phi_3-\phi_4)}\rangle
\end{eqnarray}
\cite{Borghini:2000sa,Borghini:2001vi}. 
The four-angle term depends of the four particle correlation function $\rho_4(1,2,3,4)\equiv
\rho_4(\mathbf{p}_1, \mathbf{p}_2,\mathbf{p}_3, \mathbf{p}_4)$. In systems with large numbers of
particles, multi-particle correlations are dominated by two-particle correlations. We therefore
write  
\begin{eqnarray}\label{eq:rho4}
\rho_4(1,2,3,4)
&=& \rho_1(1)\rho_1(2) \rho_1(3)\rho_1(4)\nonumber\\
& &{} + \rho_1(1)\rho_1(2)r(3,4) +\ldots \nonumber\\
& &{} + r(1,2) r(3,4) + \ldots  
\end{eqnarray}
where $\rho_1(1) \equiv  \rho_1(\mathbf{p}_1)$, etc., and the ellipses represents the distinct
permutations of the momenta. Computing $\langle \cos{n(\phi_1+\phi_2-\phi_3-\phi_4)}\rangle$, one
finds that the second and third lines in (\ref{eq:rho4}) contribute respectively terms of
order $\langle v_n\rangle^2 \sigma_n^2$ and $\sigma_n^4$ that precisely cancel the $\sigma_n$
contributions from $v_n\{2\}^4$. This cancellation is by design  -- it dictates the form of
(\ref{eq:v4FluctExp}) proposed in Ref.\  \cite{Borghini:2000sa,Borghini:2001vi}. 
One finds
 \begin{eqnarray}\label{eq:v4}
v_n\{4\}  \approx  \langle v_n\rangle, 
 \end{eqnarray}
plus corrections that are in practice very small.  The fact that $v_n\{4\}$ receives no contribution
from $\sigma_n$ is usually thought of as a consequence of the Bessel-Gaussian approximation
\cite{Voloshin:2007pc}, but we see that it is true whenever two-body correlations are dominant. 
However, note that definitions of $\langle v_n\rangle$ other than (\ref{eq:vnRP}) would yield
different results. 

As an aside, we briefly discuss the corrections to (\ref{eq:v4}). Strictly speaking, calculation of
(\ref{eq:v4FluctExp}) using (\ref{eq:rho4}) yields 
 \begin{eqnarray}\label{eq:vn4corrections}
v_n\{4\}^2  \approx  \langle v_n\rangle^2 - 2\Sigma_n^2, 
 \end{eqnarray}
 where 
\begin{equation}\label{eq:BigSigDef}
    \Sigma_n^2 =  
       \int\! d\mathbf{p}_{1}d\mathbf{p}_{2}\,
     \frac{r(\mathbf{p}_{1},\mathbf{p}_{2})}{\langle N(N-1)\rangle}
    \cos{2n(\Phi-\psi_{{}_{RP}})}
 \end{equation}
 for $\Phi = (\phi_1 + \phi_2)/2$ the average angle of the correlated pairs. Further corrections of
higher order in $1/N$ are discussed in appendix \ref{appendA}. Both $\Sigma_n$ and $\sigma_n$ are of
the same order in $1/N$, but one expects $\Sigma_n$ to be smaller than $\sigma_n$ because it effectively is a
higher harmonic, of order $2n$ rather than $n$ \cite{Borghini:2000sa}. Indeed, we find the
contribution of $\Sigma_n$ to $v_n\{4\}$ smaller than $1.2\%$ for $n= 2$ for peripheral collisions
in our model. Note that $\Sigma_n=0$ in central collisions. We comment that $\Sigma_n$ represents
fluctuations of harmonics of $\Phi$, which are important in studying the Chiral Magnetic Effect
\cite{Voloshin:2011mx}.
%
%
\begin{figure}
\includegraphics[scale=0.45]{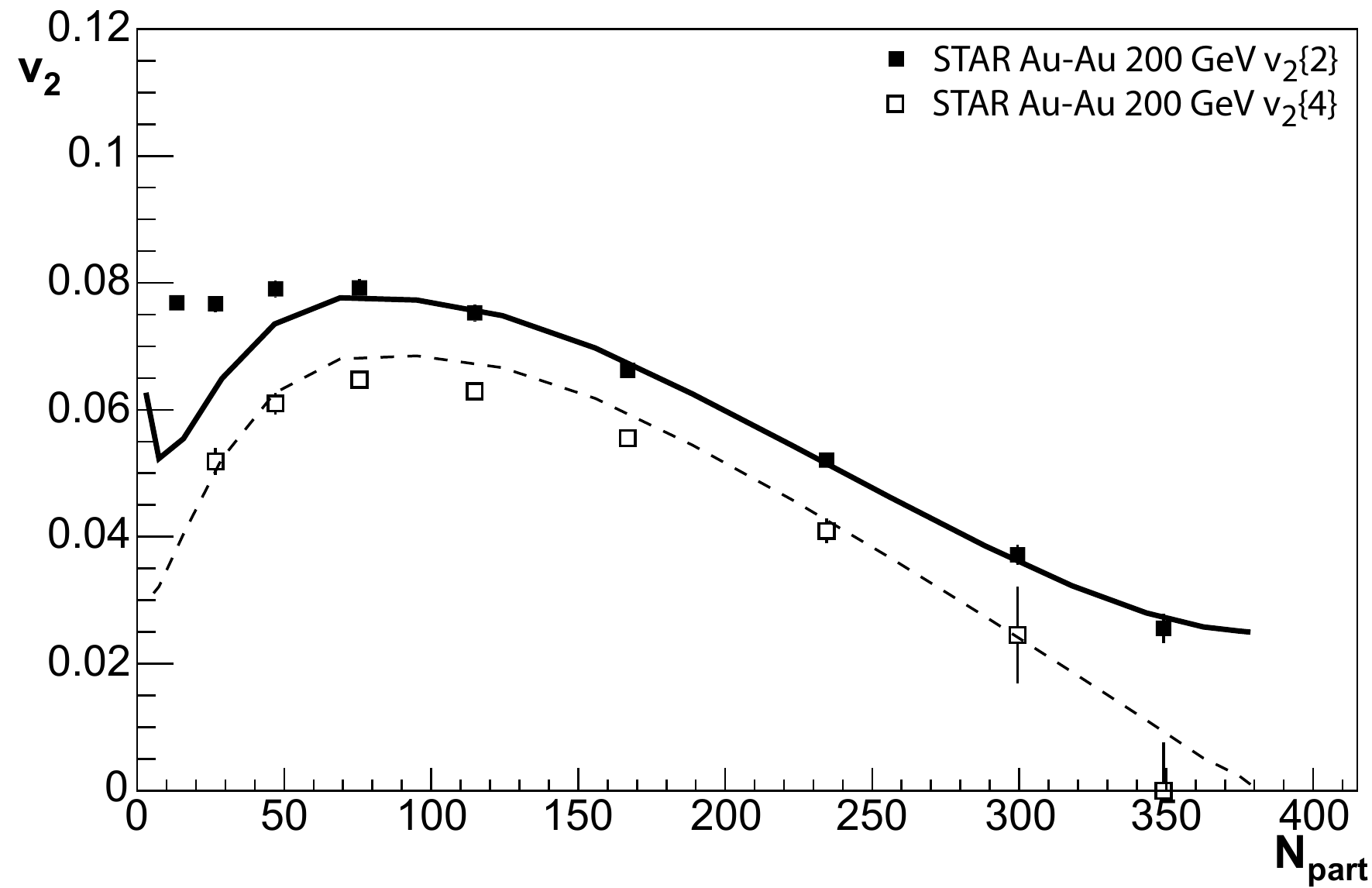}%
\caption{\label{fig:STARv2}Measured $v_2\{2\}$ and $v_2\{4\}$  from STAR \cite{Pruthi:2011eq} compared to calculations. Computed 
$v_2\{2\}$ (solid) uses (\ref{eq:vnsigma}) and (\ref{eq:sigmaDef}), while $v_2\{4\}$ (dashed) uses (\ref{eq:vnRP}), (\ref{eq:v4}), and (\ref{eq:CooperFrye}). The difference between the calculated curves is due to flow fluctuations.}
\end{figure}

Our aim is to understand the contribution of two-particle correlations to flow fluctuation measurements. 
If the only correlations are due to the reaction plane then $v_n\{2\} = v_n\{4\}$. 
This requires that no other sources of correlation exist, i.e., $r(\mathbf{p}_1, \mathbf{p}_2)=0$,
allowing $\rho_2(\mathbf{p}_1, \mathbf{p}_2)$ to factorize. However, experiments measure differences between $v_n\{2\}$ and $v_n\{4\}$ for several
harmonic orders, $n$ \cite{Pruthi:2011eq,Sorensen:2011fb,ALICE:2011ab,BilandzicThesis}. For example,
Fig.\ref{fig:STARv2} shows $v_2\{2\}$ and $v_2\{4\}$ as measured by the STAR experiment in 200
GeV Au+Au collisions \cite{Pruthi:2011eq}. One can quantify this difference in terms of
$\sigma_n$ by substituting the square of (\ref{eq:v4}) into (\ref{eq:vnsigma}), to obtain
%
%
\be
\sigma_{n}^2 = \frac{v_n\{2\}^2 - v_n\{4\}^2}{2}.
\label{eq:sigvn}
\ee
Measurements with rapidity separations should suppress contributions to $\sigma_n$ from jets,
resonance decays, and other short range non-flow correlations, but do not explain the difference
between $v_n\{2\}$ and $v_n\{4\}$, suggesting that $\la v_n^2\ra$-factorization is not a signature
of flow or flow fluctuations.

An alternative measure of flow fluctuations can be constructed using event-by-event harmonics  \cite{Mrowczynski:2002bw,Voloshin:2007pc,Ollitrault:2009ie}. If each event produces a set of harmonics $\{{\hat v}_{n}\}$, then one can compute the variance ${\hat\sigma}_{n}^2 = \langle {\hat v}_{n}^2\rangle - \langle {\hat v}_{n}\rangle^2$ for an ensemble of such events. It is then reasonable to ask how our $\sigma_n$ compares to this variance.
Taking the mean square of ${\hat v}_{n}$ to be $\langle {\hat v}_{n}^2\rangle = \la (\sum_{i}\cos n(\phi_i-\Psi_{{}_{RP}}))^2\ra/\la N\ra^2$,  we find
\be
{\hat\sigma}_n^2 =  {\hat\sigma}_{\rm stat}^2 +\sigma_n^2 + \Sigma_n^2/2,
\label{eq:ensemble}
\ee
where $\sigma_n$ and $\Sigma_n$ are given by (\ref{eq:sigmaDef}) and (\ref{eq:BigSigDef}), and we keep only leading order in $N$. We define the statistical variance  
\be
{\hat\sigma}_{\rm stat}^2 = \frac{1+\langle v_{2n}\rangle}{2\langle N\rangle}.
\label{eq:stat}
\ee
These fluctuations arise because a finite number of particles are sampled in each event. In contrast, the dynamical fluctuations $\sigma_n$ and $\Sigma_n$ are caused by the correlations between particles \cite{Pruneau:2002yf}. To an excellent approximation the $v_{2n}$ contribution to (\ref{eq:stat}) and the $\Sigma_n^2$ contribution to (\ref{eq:ensemble}) can be neglected.

We see that the event-wise variance of ${\hat v}_{n}$ is very different from the quantity (\ref{eq:sigvn}) that characterizes the difference between $v_n\{2\}$ and $v_n\{4\}$.  The statistical contribution $\hat\sigma_{\rm stat}$ can be comparable in magnitude to $\sigma_n$. 
We point out that the quantity $\hat\sigma_n$ may be measured directly from event-by-event harmonic coefficients. 

It is instructive to compare our results to the model of Ref.\ \cite{Voloshin:2007pc,Ollitrault:2009ie} in which flow fluctuations are exclusively due to the event-wise variation of the global geometry. In the purely geometric interpretation, one defines an eccentricity for each event $\hat\epsilon_{n}$. If one assumes  that the relation between $\hat\epsilon_{n}$ and the resulting anisotropy of the fluid flow ${\hat v}_n$ is approximately deterministic, then fluctuations of the ratio ${\hat v}_{n}/\hat\epsilon_{n}$ are negligible. The geometric contribution to the variance is then $\hat\sigma_n^2/\langle v_n\rangle^2 =(\langle \hat\epsilon_n^2\rangle - \langle \hat\epsilon_n\rangle^2)/\langle \hat\epsilon_n\rangle^2$. We mention that Glauber model estimates of the relevant eccentricities are consistent with data for $n=2$, where the correlation of ${\hat v}_{n}$ with an event-wise eccentricity is highly plausible and well established \cite{deSouza:2011rp}. However, the applicability of such an approach for $n\neq 2$ is less clear. 

We comment that the purely geometrical description of flow fluctuations in Ref.\ \cite{Ollitrault:2009ie} requires $\la {\hat v}_n\ra \gg {\hat \sigma}_n$ in addition to ${\hat v}_{n}\propto \hat\epsilon_{n}$. Neither our general discussion here nor our specific calculations in Sec.\ \ref{sec:BW} require these assumptions.  In Sec.\ \ref{sec:BW}, we include geometric fluctuations in our estimates of flow fluctuations through the
distribution of flux tube sources $\rho_{{}_{FT}}$. Correlations are then modified by the local transverse expansion
of the system. 
%
%
%
%
\section{\label{sec:loco}Local Correlations}
To understand the source of flow fluctuations, we recall some lessons from general fluctuation studies \cite{Gazdzicki:1992ri,Mrowczynski:1998vt,Pruneau:2002yf,Westfall:2008zz}. Such studies typically focus on the variation of bulk observables such as multiplicity or transverse momentum within an ensemble of collisions.  We draw most heavily from the study of $p_t$ fluctuations in Ref.\  \cite{Gavin:2011gr}, to which the present work is essentially a sequel. 

We measure $p_t$ fluctuations using the covariance $\langle \delta p_{t1}\delta p_{t2}\rangle$, where $\delta p_{ti} = p_{ti}-\langle p_t\rangle$ is the deviation of each $p_t$ from the average. Pairs  in which both particles have higher than average momentum add to  $\langle \delta p_{t1}\delta p_{t2}\rangle$. Lower-than-average pairs also add to the covariance, while high/low pairs subtract from it. In global equilibrium $\langle \delta p_{t1}\delta p_{t2}\rangle\equiv 0$. The presence of hot spots makes $\langle \delta p_{t1}\delta p_{t2}\rangle > 0$  (as would cold spots) \cite{Gavin:2003cb}. Motion of the sources further enhances this quantity \cite{Voloshin:2003ud}. It follows that both jets and flow add to the $p_t$ covariance. 

Moving hot spots in the fluctuating system are sources of local correlations. We distinguish such correlations from those due to the global event-wise variation of the shape and size of the collision volume.  We write the $p_t$ covariance as 
\begin{equation}\label{eq:Dynamic}
    \langle \delta p_{t1}\delta p_{t2}\rangle =
    \int\! d\mathbf{p}_{1}d\mathbf{p}_{2}\,
    \frac{r(\mathbf{p}_{1},\mathbf{p}_{2})}{\langle N(N-1)\rangle}
    \delta p_{t1} \delta p_{t2}.
\end{equation}
In \cite{Gavin:2011gr}  we demonstrate that the $p_t$ covariance is independent of the global spatial anisotropy, because (\ref{eq:Dynamic}) is integrated over azimuthal angle. It is also independent of size fluctuations by construction \cite{Voloshin:1999yf}. 

To see that flow fluctuations are also primarily driven by local correlations, observe that (\ref{eq:Dynamic}) has the same form as (\ref{eq:sigmaDef}) with $\delta p_{t1} \delta p_{t2}$ replaced by $\cos n\Delta\phi$.  As the $p_t$ covariance probes correlations in momentum, so $\sigma_n$ measures correlations of the angular separation $\Delta\phi$ of pairs.  

This analogy makes physical sense because of the common influence of hot spots on $\Delta\phi$ and $p_t$. Flow pushes particles from a source into a particular opening angle, depending on the source position and the local fluid velocity. The same flow velocity boosts the $p_t$ of these particles. Therefore multiple random hot spots give rise to both flow and $p_t$ fluctuations.  
Based on this analogy, we expect $\sigma_n$ in central collisions to be non-zero for all orders $n$, while $\langle v_n\rangle$ defined by (\ref{eq:vnRP}) vanishes.  Furthermore, it is not likely that $\sigma_n$ is as strongly influenced by the global geometry as the mean $\langle v_n\rangle$.  

In the next sections we discuss flow fluctuations due to early time local spatial correlations contributing to $r(\mathbf{p}_1, \mathbf{p}_2)$. We employ a Glasma-flux tube model in which particles are initially correlated at the point of production and modified by later stage transverse expansion. These flux tubes produce our `hot spots'.  Transverse expansion is modeled with a blast wave scenario which inherits anisotropy from the average eccentricity $\la \varepsilon_{_{RP}}\ra$.

%
%
\section{\label{sec:glasma}Source of Correlations}
Flow transforms early-time spatial correlations into observable momentum correlations. Two-particle correlations probe the initial state average pair distribution $n_2(\mathbf{x}_1,\mathbf{x}_2)$, which encompasses both genuine pair correlation as well as correlations due to fluctuating event eccentricities, shapes, and orientations (with respect to the reaction plane). Define the initial state spatial
correlation function
%
%
\be
c(\mathbf{x}_1, \mathbf{x}_2) = 
n_2(\mathbf{x}_1,\mathbf{x}_2) - n_1(\mathbf{x}_1)n_1(\mathbf{x}_2),
\label{eq:CorrDef}
\ee
where $n_1(\mathbf{x}_i)$ is the particle density. In the absence of genuine two-particle correlations the pair distribution factors into a product of one body distributions, so that the correlation function vanishes. Correlations come about because the position and number of the colliding nucleons fluctuates in individual events. In general, these fluctuations result in high-density `hot spots' of varying size distributed throughout the collision volume. It is more likely to find pairs of particles near hot spots. Averaging over an ensemble of events therefore yields a nonzero $c(\mathbf{x}_1, \mathbf{x}_2)$. We define the correlation strength $\R$ as $\int c(\mathbf{x}_1, \mathbf{x}_2) d^3\mathbf{x}_1d^3\mathbf{x}_2 \equiv \la N\ra^2\R$. 

In Ref.\cite{Gavin:2011gr}  we construct a correlation function (\ref{eq:CorrDef}) to describe a specific early-time scenario based on a Glasma theory. In that picture parton production arises from the fragmentation of longitudinal color fields created at the moment of the collision. It is helpful to think of the longitudinal fields as flux tubes of transverse size $\sim Q_s^{-1}$, where $Q_s$ is the saturation scale. The saturation scale depends on many collision variables including the density of participant nucleons and the collision energy \cite{Kharzeev:2000ph,Kharzeev:2004if}. 

Particle production then proceeds via a system of flux tubes distributed over the collision volume. Given that the transverse area of a flux tube is much smaller than the transverse collision area and taking particles emerging from the same tube as correlated, we write 
%
%
\be
c(\mathbf{x}_1, \mathbf{x}_2)
 = {\la N\ra}^2\R\,\delta(\mathbf{r}_t) \rho_{_{FT}} (\mathbf{R}_t).
\label{eq:CorrFunc}
\ee
The spatial coordinates are relative, $\mathbf{r}_t = \mathbf{r}_{t1} - \mathbf{r}_{t2}$, and
average, $\mathbf{R}_t = (\mathbf{r}_{t1} + \mathbf{r}_{t2})/2$, transverse elliptical coordinates.
The delta function enforces the correlation due to the common point of production, and $\rho_{_{FT}}
(\mathbf{R}_t)$ is the average transverse distribution of flux tubes
%
%
\be
\rho_{_{FT}}(\mathbf{R}_t) \approx  \frac{2}{\pi R^2_A}  \lp 1 -\frac{R_t^2}{R_A^2}\rp.
\label{eq:TubeDis}
\ee
Here the shape of (\ref{eq:TubeDis}) resembles the nuclear thickness function and $\pi R_A^2$ is the
area of the overlap region. The distribution (\ref{eq:TubeDis}) represents an average over all
possible shapes, which we have taken to be a simple ellipse.

To compute the correlation strength $\cal R$, we imagine each event produces $K$ flux tubes with transverse size $\sim Q_s^{-2}$. In the saturation regime $K$ is proportional to the transverse area $R_A^2$ divided by the area per
flux tube, $Q_s^{-2}$ \cite{Kharzeev:2000ph}. Allowing $K$ to fluctuate from event to event with
average $\la K \ra$, we calculate in Ref.\cite{Gavin:2008ev} that
%
\be
\R  =\frac{\la N^2 \ra -\la N \ra^2 -\la N \ra}{\la N \ra^2}
\propto \langle K\rangle^{-1}.
\label{eq:Rdef}
\ee
Each Glasma flux tube yields an average multiplicity of  $\sim \alpha_s^{-1}(Q_s)$ gluons
and as in Ref.\cite{Kharzeev:2000ph}, the number of gluons in a rapidity interval $\Delta y$ is then
%
\be
\langle N\rangle = ({{dN}/{dy}})\Delta y  \sim {\alpha_s}^{-1}(Q_s)\langle K\rangle.
\label{eq:Nscale}
\ee
Finally, the Glasma correlation scale,
%
\be
\R{{dN}/{dy}} = \kappa {\alpha_s^{-1}}(Q_s^2),
\label{eq:CGCscale}
\ee
follows from the multiplication of (\ref{eq:Rdef}) and (\ref{eq:Nscale}). Notice
(\ref{eq:CGCscale}) is dimensionless and depends only on the saturation scale, \Qs, which can be
calculated from first principles. 
Measurements of the ridge at various beam energies, target masses, and centralities fix the
dimensionless coefficient $\kappa$ and are in excellent accord with the
leading-order dependence  \cite{Gavin:2008ev,Moschelli:2009tg}. 

In the next section, we will see how transverse expansion modifies Glasma correlations
(\ref{eq:CorrFunc}), but it is important to note that (\ref{eq:CGCscale}) significantly impacts
the centrality dependence and determines virtually all of the energy dependence.
%
%
\section{\label{sec:BW}Flow Fluctuations from Glasma and Transverse Expansion}
Transverse collective expansion gives rise to momentum space correlations in two ways. First, the event geometry influences global correlations; collective flow emerges with respect to a reaction plane. Second, the local fluctuations, that together determine the global shape, individually induce local correlations; flow, on average, modifies the momenta from all partons emerging from the same fluctuation in a common way.

To understand how flow modifies local correlations we first model the average anisotropic expansion
of the system with a blast wave model
\cite{Gavin:2008ev,Moschelli:2009tg,Schnedermann:1993ws,Retiere:2003kf,Huovinen:2001cy}. 
As in Refs.\cite{Gavin:2008ev,Moschelli:2009tg} the invariant single-particle distribution, 
%
%
\be
\rho_1\lp\mathbf{p}\rp\equiv\frac{dN}{dyd^2p_t} =
\int f\lp\mathbf{x},\mathbf{p}\rp d\Gamma,
\label{eq:CooperFrye}
\ee
is defined as the particle flux through a surface, $\sigma$ \cite{Cooper:1974mv}. At constant proper
time, $\tau_F$, the flux element is $d\Gamma=p^\mu d\sigma_\mu = \tau_F m_t \cosh(y-\eta)d\eta
d^2 r$, where $\eta = (1/2)\ln((t+z)/(t-z))$ is the spatial rapidity.
Here $f(\mathbf{x},\mathbf{p}) =  {(2\pi)^{-3}}\exp\{-u^\mu p_\mu/T\}$ is the Boltzmann phase-space
density for a temperature $T$ and fluid four-velocity $u_\mu$. We follow
Ref.\cite{Schnedermann:1993ws} and write the four velocity of the
longitudinal-boost invariant blast wave as $u_\mu = \gamma_t (\cosh\eta, \mathbf{v}_t, \sinh\eta)$,
with $\mathbf{v}_t$, the transverse velocity and $\gamma_t = (1-v_t^2)^{-1/2}$. The phase space
density is then $f \propto \exp\{-\gamma_t m_t \cosh(y-\eta)/T\}  \exp\{\gamma_t \mathbf{v}_t \cdot
\mathbf{p}_t/T\}$.

To account for the initial elliptical shape and anisotropic velocity we require one additional
parameter, the (average) event eccentricity, $\varepsilon$. The anisotropic transverse flow
velocity, 
%
%
\be
\gamma_t \mathbf{v}_t = \lambda\lp\varepsilon_x \mathbf{x} + \varepsilon_y \mathbf{y}\rp,
\label{eq:anisoV}
\ee
mimics the geometrical parton density gradients. To account for the larger pressure in the
direction of the reaction plane ($x$-direction in (\ref{eq:anisoV})), we take $\varepsilon_{x,y} =
\sqrt{1 \pm \varepsilon}$. Rewriting Eq.(\ref{eq:anisoV}) as $(\gamma_t v_t)^2 =
\lambda^2 r^2(1+\varepsilon\cos(2\phi))$, one can see how this choice of flow velocity results in an
asymmetric flow velocity peaked in the direction of the reaction plane. Note, $r$ is now the
elliptical polar radius and depends on $\varepsilon$ and the spatial angle.

In this formulation, particles passing through the freeze out surface have a velocity, $v_s$, that
is connected to the position of production, thus the transverse velocity is constrained via
$\gamma_t^2v_s^2 =\lambda^2R_{surface}^2$. As in \cite{Gavin:2011gr},
we take the blast wave parameters $v_s$ and $T$ from experiment \cite{Kiyomichi:2005zz} and
parameterize $\varepsilon$ to fit $v_2$.
However, in Ref.\cite{Gavin:2011gr}, we did not include the impact of fluctuations on the
calculation of $v_n$-like correlations - a focus in this work. Hence, in
Ref.\cite{Gavin:2011gr}, $\varepsilon$ is chosen to fit $v_2\{2\}$. In this work we identify
$\la v_2\ra_{_{BW}}$ with purely reaction plane angular correlations, more suitably comparable to
$v_2\{4\}$. In Fig.\ref{fig:STARv2}, the dashed line shows this comparison where the open squares 
are the STAR Au-Au 200 GeV $v_2\{4\}$ measurement \cite{Pruthi:2011eq}.

Now that our blast wave formalism is tuned to reproduce the average transverse expansion in heavy
ion collisions, we are in a position to calculate the momentum space two-particle correlation
function. As discussed by other authors, see e.g.
\cite{Borghini:2007ku,Pruneau:2006gj,Danielewicz:1987in}, the two-particle joint probability
distribution (\ref{eq:pairDensity}) which we rewrite as 
%
%
\be
r(\mathbf{p}_1, \mathbf{p}_2) \equiv
\rho_2 (\mathbf{p}_1, \mathbf{p}_2) - 
\rho_1(\mathbf{p}_1)\rho_1(\mathbf{p}_2) 
\label{eq:MomCorr0}
\ee
to highlight the fact that any genuine correlations, including those from local initial state
fluctuations (\ref{eq:CorrDef}), break the factorization
$\rho_2(\mathbf{p}_1,\mathbf{p}_2)\rightarrow\rho_1(\mathbf{p}_1)\rho_1(\mathbf{p}_2)$. 

The spatial correlation function, (\ref{eq:CorrFunc}), embodies the correspondence between the two
particle joint probability distribution and transverse expansion. 
Intrinsically correlated partons originate from the same transverse position and experience, on
average, the same transverse momentum modulation from flow. 
Generalizing (\ref{eq:CooperFrye}), we represent the genuine two-particle correlation as
%
%
\be
r(\mathbf{p}_1, \mathbf{p}_2) =
\!\!\int c(\mathbf{x}_1, \mathbf{x}_2) 
f(\mathbf{x}_1,\mathbf{p}_1)
f(\mathbf{x}_2,\mathbf{p}_2)
d\Gamma_1d\Gamma_2.
\label{eq:MomCorr}
\ee
Notice that full integration of (\ref{eq:MomCorr}) yields $\la N\ra^2\R$ as does (\ref{eq:CorrDef}),
tying azimuthal correlations to multiplicity and transverse momentum fluctuations
\cite{Gavin:2011gr}.

To compute flow fluctuations using this correlation function, we combine (\ref{eq:sigmaDef}) with (\ref{eq:CorrFunc}) and (\ref{eq:MomCorr}). We  then use (\ref{eq:vnsigma}), (\ref{eq:v4}), and (\ref{eq:sigmaDef}) to calculate $v_2\{2\}$ and $v_2\{4\}$ in Fig.\ref{fig:STARv2}, where the flow coefficient relative to the reaction plane is  $\langle v_n\rangle = \int \rho_1(\mathbf{p})\cos n(\phi-\Psi_{_{RP}})d\mathbf{p}$. 
It is significant that $\sigma_n^2 \propto \R$; this is also true for the long range contribution to other fluctuation quantities \cite{Gavin:2011gr}. 

The influence of event shapes and eccentricities enters (\ref{eq:MomCorr}) in two ways. First, the correlation function $c(\mathbf{x}_1, \mathbf{x}_2)$ includes a probability distribution of sources (\ref{eq:TubeDis}) that implicitly accounts for event-wise variation of the volume and eccentricity. Integration over spatial coordinates effectively represents an average over all possible event shapes. Second, the azimuthal distribution of particles from each source is modified by flow. A pair emerging from a source that experiences a greater push will have a narrower opening angle. The magnitude of the push follows (\ref{eq:anisoV}) and depends on position. 

It is important to note that any radial expansion will induce correlations contributing to all orders of $v_n$. Particular choices of the flow velocity affect their relative contributions, i.e. an elliptical flow velocity profile induces a large $v_2$ and a triangular flow velocity profile would induce a large $v_3$. Equation (\ref{eq:anisoV}) is chosen to mimic the average (elliptical) flow profile.

The long range behavior introduced by Glasma correlations provides key centrality and collision energy dependence on the saturation scale \Qs~through $\R$ in  (\ref{eq:CGCscale}). 
Although, the blast wave parameters
$T$, $v_s$, and $\varepsilon$ also provide significant centrality dependence, their change with
collision energy is minimal. 
The average $v_s$ and freeze out temperature in 62 GeV and 200 GeV Au+Au collisions
are the same as those used in \cite{Gavin:2011gr, Gavin:2008ev,Moschelli:2009tg} and are based on an
analysis in \cite{Kiyomichi:2005zz}.  At 2.76 TeV the velocity is scaled up from the 200 GeV values
by 6\% and the temperature is scaled up by 7\% as presented in \cite{Gavin:2011gr}.
%
%
\begin{figure}
\includegraphics[scale=0.45]{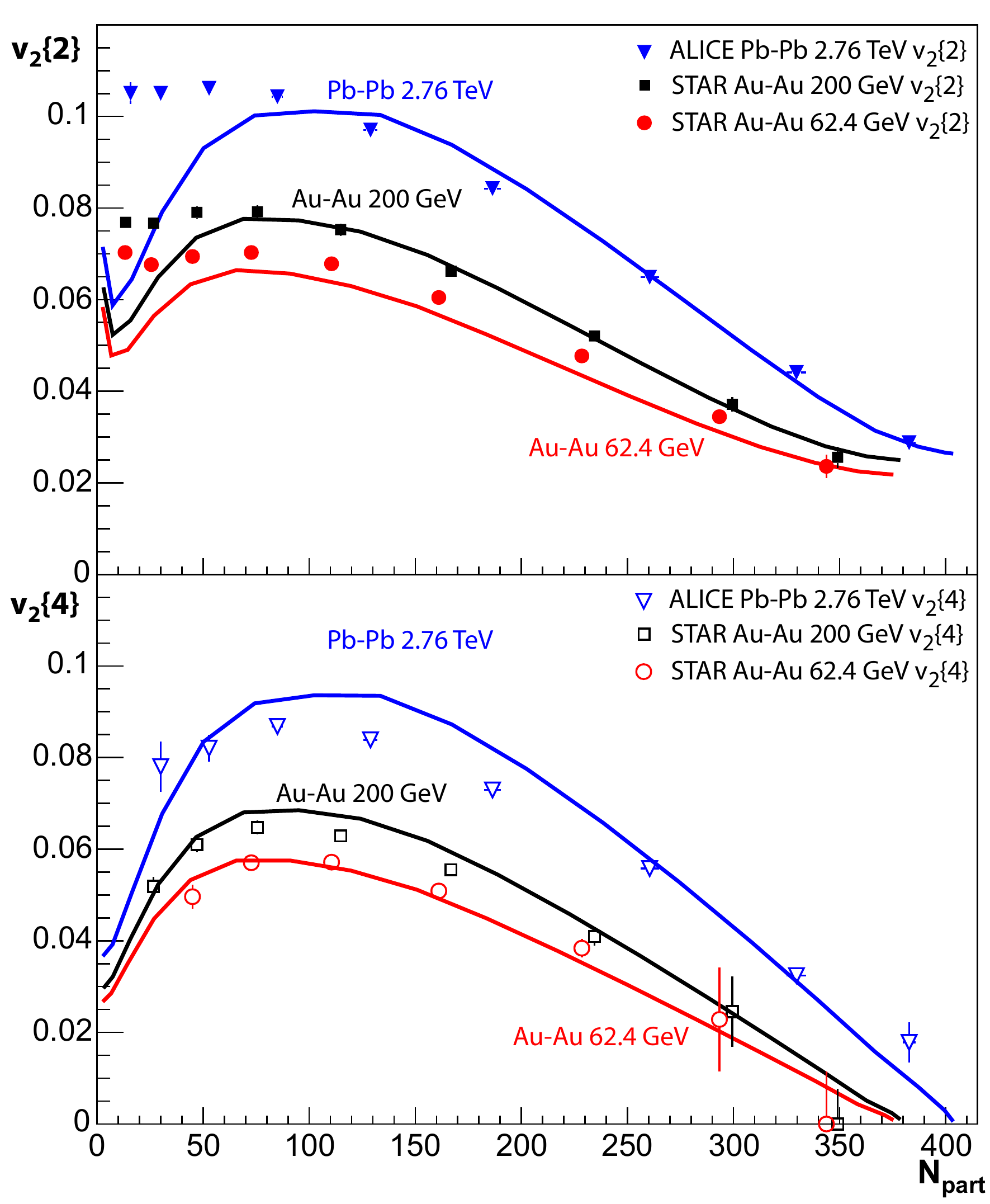}%
\caption{\label{fig:v22}Measured $v_2\{2\}$ (top)  from STAR 
\cite{Pruthi:2011eq} and ALICE \cite{ALICE:2011ab,BilandzicThesis} compared to calculations
using (\ref{eq:vnsigma}) and (\ref{eq:sigmaDef}).  Same for $v_2\{4\}$ (bottom) computed from 
(\ref{eq:vnRP}), (\ref{eq:v4}), and (\ref{eq:CooperFrye})}.
\end{figure}

\begin{figure}
\includegraphics[scale=0.45]{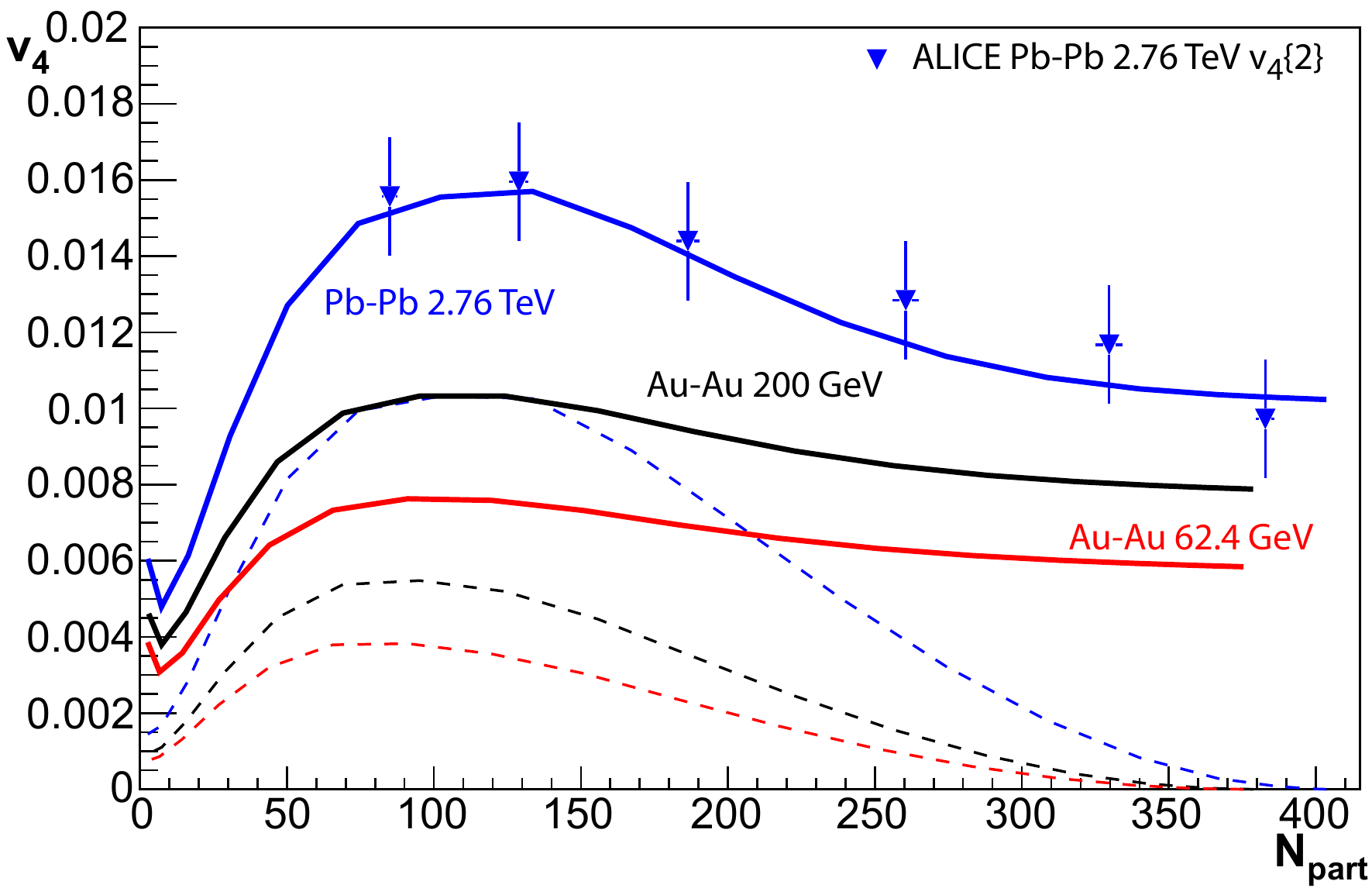}%
\caption{\label{fig:v4} Same calculation as in Fig.\ref{fig:v22} but for the fourth order flow
coefficients $v_4\{2\}$ (solid) and $v_4\{4\}$ (dashed). Data is from the ALICE collaboration \cite{ALICE:2011ab}.}
\end{figure}

In Fig.\ref{fig:v22} we show comparisons of our calculated $v_2\{2\}$ (top panel) and $v_2\{4\}$
(bottom panel) to measured data \cite{Pruthi:2011eq,ALICE:2011ab,BilandzicThesis} and find the
agreement concerning change in collision energy is quite good.
Similarly in Fig.\ref{fig:v4} we calculate $v_4\{2\}$ and  $v_4\{4\}$, 
represented as the solid and dashed lines, respectively. 
Comparisons to $v_n\{4\}$ in both figures show how the eccentricity $\varepsilon$ affects the results. Recall that $\varepsilon$ is tuned to fit $v_2$ in Au+Au 200 GeV collisions. The effect of eccentricity is to reduce the even harmonics as the collision area becomes circular and  $\varepsilon\rightarrow 0$. 
The non-zero $v_n\{2\}$ in central collisions are due to fluctuations (\ref{eq:sigvn}), which in our case are due to Glasma correlations (\ref{eq:MomCorr}). 
The effect of the change in blast wave parameters with collision energy is apparent from the
separation of the solid lines in the bottom panel of Fig.\ref{fig:v22} and the dashed lines in Fig.\ \ref{fig:v4}. The importance of the Glasma
contribution is similarly evident by the change is separation between $v_n\{4\}$ and $v_n\{2\}$ with
collision energy. The increase in growth of flow fluctuations (\ref{eq:sigmaDef}) is coupled to the
growth in the saturation scale.
%
%

STAR has also measured $v_2$ fluctuations in the form of
%
%
\be
\frac{\sigma_{v_n}}{\la v_n\ra} =
\sqrt{\frac{v_n\{2\}^2 - v_n\{4\}^2}{v_n\{2\}^2 + v_n\{4\}^2}}
\label{eq:starFano}
\ee
resembling the so-called ``coefficient of variation'' defined as the standard deviation divided by the mean. Care should be taken here since the
definition of $\sigma^2_{n}$, Eq.({\ref{eq:sigvn}), is not strictly the variance. 
In Fig.\ref{fig:fano} we compare our calculation of (\ref{eq:starFano}) to measurement.
Calculations at RHIC energies seem to agree reasonably well and we include the calculation for
Pb+Pb 2.76 GeV as a prediction.
%
%
\begin{figure}
\includegraphics[scale=0.45]{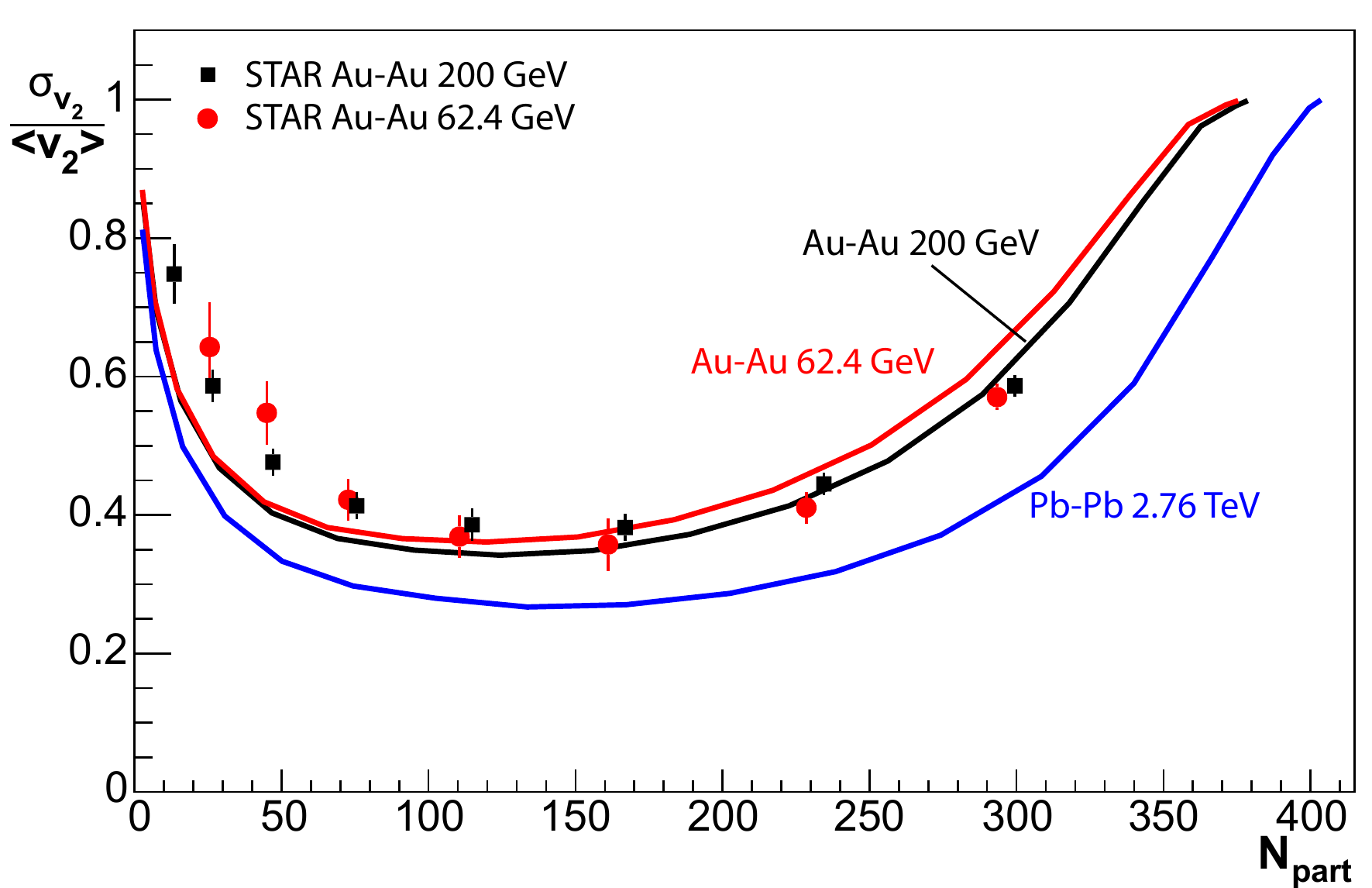}%
\caption{\label{fig:fano}Flow fluctuations in terms of the coefficient of variation for elliptic flow (\ref{eq:starFano}) compared to
STAR data \cite{Pruthi:2011eq}.}
\end{figure}

Observe that local correlations that contribute to $\sigma_{n}$ lead to fluctuations of all harmonic orders in $n$ including $v_3$. As a result, Glasma correlations generate a measurable $v_3\{2\}$ even in the absence of $v_3\{4\}$ or any triangular flow. In Fig.\ref{fig:sig_v3} we show $v_3$ fluctuations from Glasma. We further emphasize that  the energy dependence in Fig.\ref{fig:sig_v3} is in good accord with data, supporting the Glasma scaling with \Qs. The shape with centrality reflects contributions not only from Glasma, but also our parameterizations of the average reaction plane eccentricity
$\varepsilon$ and our choice the flux tube distribution $\rho_{_{FT}}$.

To use (\ref{eq:sigmaDef}) and ({\ref{eq:sigvn}) to calculate $v_3\{2\}$, we must come to grips with the fact that our
blast wave parametrization assumes $v_3\{4\}=0$. While this seems to be the case for the STAR
measurements, ALICE has measured a non-zero $v_3\{4\}$ for Pb+Pb collisions at 2.76
TeV. To correct for this possible discrepancy, we can provide an ad-hoc parameterization of  $v_3\{4\}$ and use (\ref{eq:sigmaDef}) and
({\ref{eq:sigvn}) to calculate $v_3\{2\}$. Agreement shown in Fig.\ref{fig:v3} is
reasonable. As a preferable alternative, we compare our calculated $\sigma_{n}$ to fluctuations extracted from ALICE and STAR measurements in Fig.\ref{fig:sig_v3}.
%
%
\begin{figure}
\includegraphics[scale=0.45]{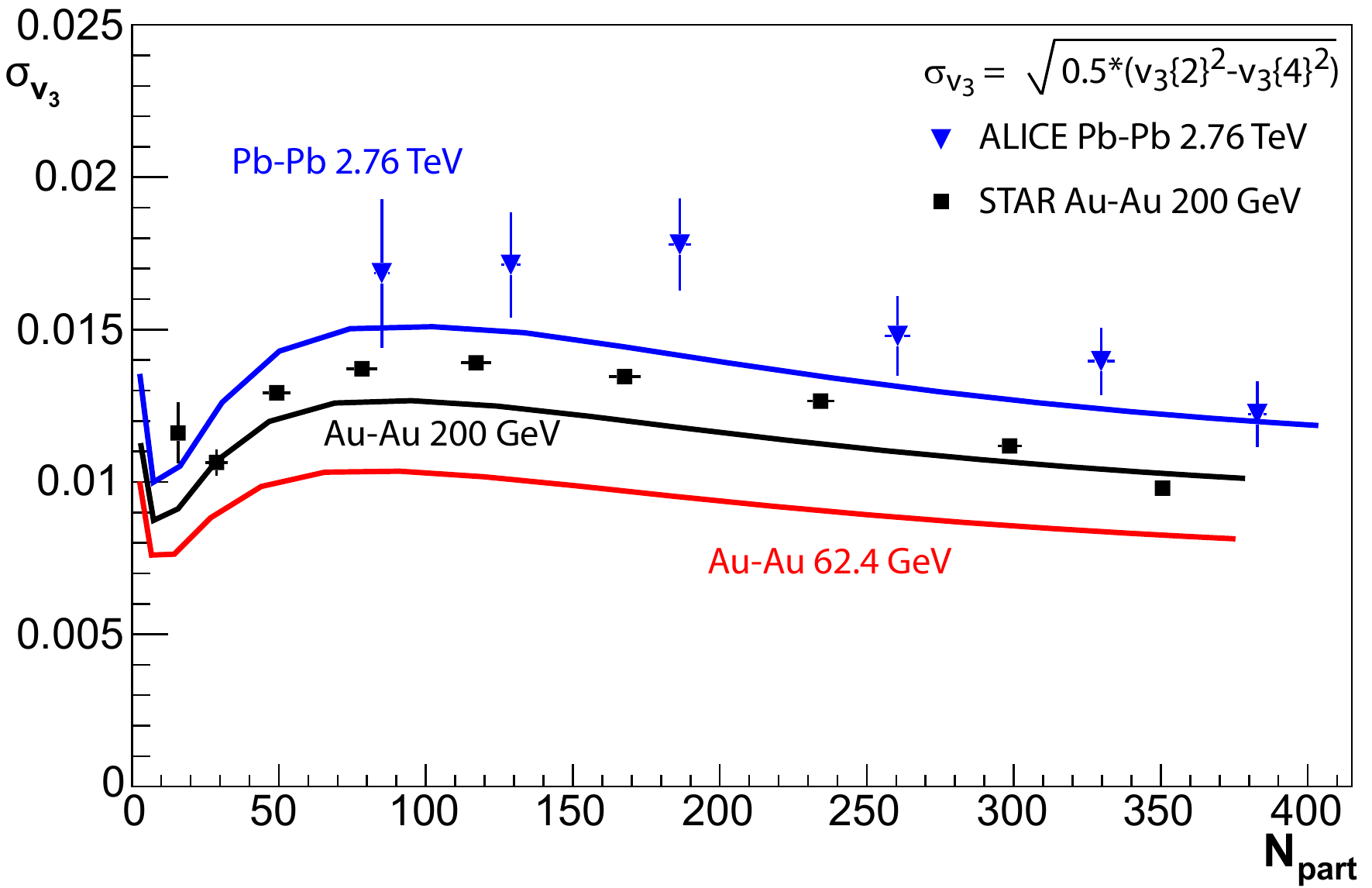}%
\caption{\label{fig:sig_v3} Triangular flow fluctuations, $\sigma_{v_3}$, calculated from
(\ref{eq:sigmaDef}) and compared to data computed from $v_3\{2\}$ and $v_3\{4\}$ measurements by
the STAR and ALICE collaborations \cite{Sorensen:2011fb,ALICE:2011ab}.}
\end{figure}
%
%
\begin{figure}
\includegraphics[scale=0.45]{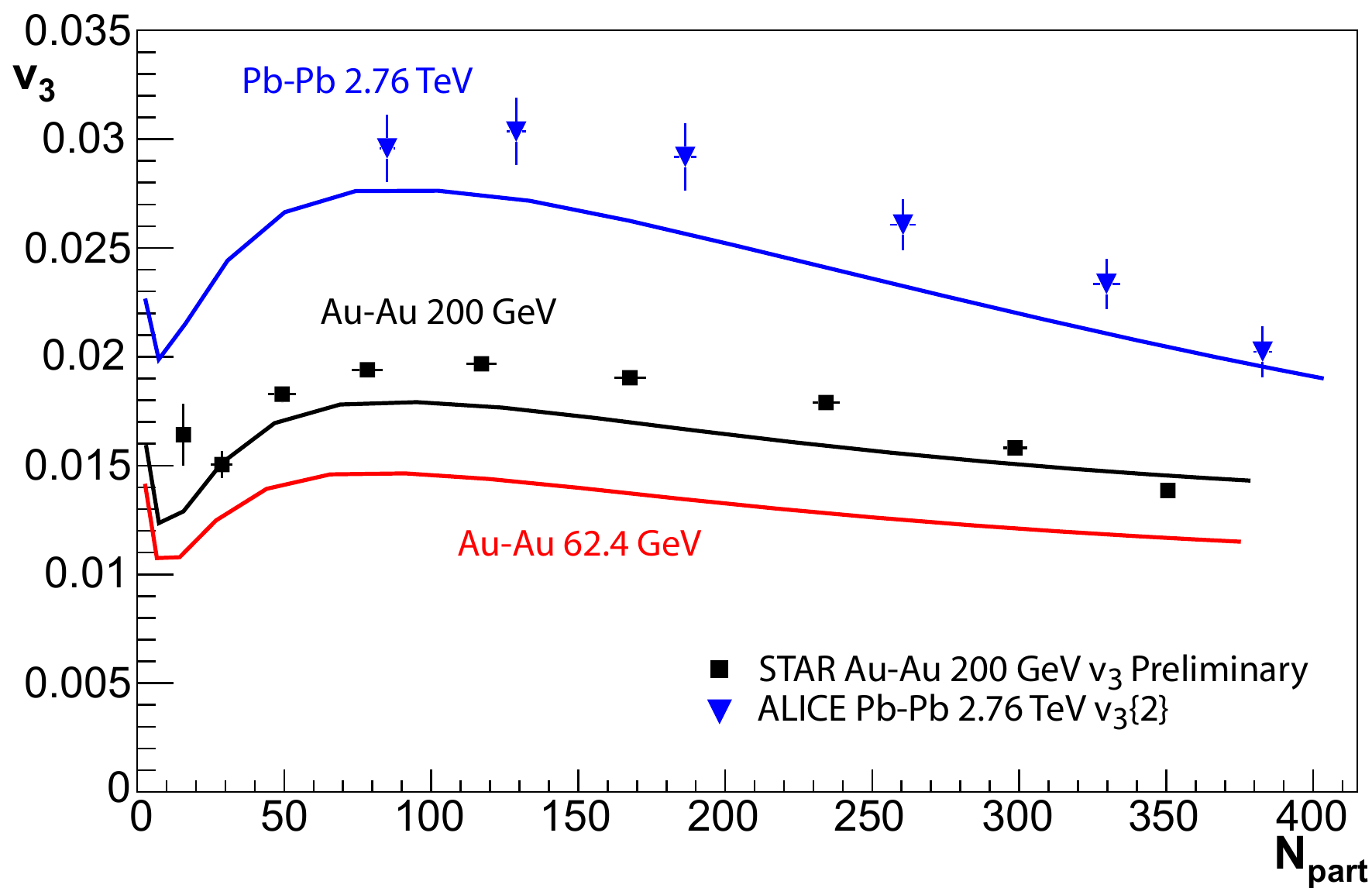}%
\caption{\label{fig:v3}The triangular flow coefficient, $v_3\{2\}$ compared to
STAR and ALICE data \cite{Sorensen:2011fb,ALICE:2011ab}.}
\end{figure}

%
%
\section{\label{sec:ridge}The Ridge}

Early analyses of the soft or untriggered ridge, including our own, focus on the idea that the
ridge is composed of correlations in excess of momentum conservation and elliptic flow.
The novel observation of Alver and Roland that correlations resulting from geometrical fluctuations
result in odd flow harmonics such as triangular flow, $v_3$, suggests that $v_3$, together with flow
harmonics of all higher orders, explains this excess \cite{Alver:2010gr}. In this paper we have
studied how angular correlations tied to common production points influence two-particle correlation
measurements of flow harmonics $v_n\{2\}$ through flow fluctuations $\sigma_n^2$. Most notably, as
discussed in Sec.\ref{sec:BW}, we find significant contributions to $v_3\{2\}$ even in the absence
of triangular flow. In fact, the same correlations contributing to flow fluctuations were initially
proposed as explanations of the ridge
\cite{Gavin:2008ev,Moschelli:2009tg,Dumitru:2008wn,Dusling:2009ar,Voloshin:2003ud,Pruneau:2007ua,
Lindenbaum:2007ui,Peitzmann:2009vj,Takahashi:2009na,Andrade:2010xy,Werner:2010aa,Sharma:2011nj},
although their role in reproducing the full $\Delta\phi$ azimuthal structure was not understood.

Flow and its fluctuations both contribute to the two-particle correlation landscape. The pair distribution is 
%
%
\be
\label{eq:rho2fourier}
\rho_2(\Delta\phi) = \rho_{\rm ref}
\lp 1+2\sum\limits_{n=1}^{\infty}\la v_n\ra^2\cos n\Delta\phi\rp
+ r(\Delta\phi), 
\ee
where $\rho_{\rm ref}$ is the experimental mixed-event background. Observe that the Fourier coefficients of (\ref{eq:rho2fourier}) reproduce (\ref{eq:vnsigma}). We take $\rho_{\rm ref}=\frac{1}{2\pi}\int\overline{\rho_1}\int\overline{\rho_1}$ where the overbar
indicates an event plane average, to mimick the experimental mixed event technique for constructing
$\rho_{\rm ref}$. The quantity $\Delta\rho(\Delta\phi) = \rho_2(\Delta\phi) - \rho_{\rm ref}$ characterizes
the angular correlations. STAR measures the ratio
%
%
\be
  \frac{\Delta\rho(\Delta\phi)}{\sqrt{\rho_{\rm ref}}}
  = \frac{1}{2\pi}\frac{dN}{d\eta}\lp 2\sum\limits_{n=1}^{\infty} \la v_n \ra^2 \cos n\Delta\phi
  + \frac{r(\Delta\phi)}{\rho_{\rm ref}}\rp,
\ee
where the $\la v_n\ra$ terms follow from (\ref{eq:vnRP}). 

STAR performs a multi-parameter fit to this distribution to separate the ridge peak from the elliptic flow contributions (along with momentum conservation and HBT contributions , which we omit for clarity).  They report  a flow-subtracted ridge amplitude $(\Delta\rho(\Delta\phi,\Delta\eta)/\sqrt{\rho_{\rm ref}})|_{FS}$ on the near side, centered at $\Delta\phi=\Delta\eta=0$ \cite{Daugherity:2006hz,Daugherity:2008zz,Daugherity:2008su}. We use (\ref{eq:CGCscale}) and (\ref{eq:MomCorr}) to calculate the flow-subtracted 
ridge amplitude
%
%
\be
\left. \frac{\Delta\rho}{\sqrt{\rho_{\rm ref}}}\right|_{FS} = \R \frac{dN}{dy}F(\Delta\phi),
\label{eq:CGCdRho}
\ee
where $F(\Delta\phi)\propto r(\Delta\phi)$ is the angular correlation function normalized so that  $\int F(\Delta\phi) d\Delta\phi =1$. The energy and centrality independent scale constant $\kappa$ in  (\ref{eq:CGCscale})  is fixed by Au-Au 200 GeV data as in \cite{Gavin:2008ev,Moschelli:2009tg}. 

The blast wave parameters have little energy dependence and the Glasma factor (\ref{eq:CGCscale})
allows for strong agreement with the 62 GeV data without adjustment of $\kappa$. Additionally, the
ALICE collaboration has measured (\ref{eq:CGCdRho}) in Pb+Pb collisions at $\sqrt{s}$ =2.76 TeV
\cite{collaboration:2011um}, providing a further test of the CGC-Glasma energy dependence. 
We confront the ALICE data in the same way, without adjusting $\kappa$, however the ALICE
\cite{collaboration:2011um} measurement procedure differs slightly than that from STAR
\cite{Daugherity:2008su}.
This difference in fitting procedures is mostly insignificant, especially in central collisions,
but not necessarily so in peripheral collisions to which we normalize our calculation (at 200 GeV).
To adjust for this possible effect, we multiply (\ref{eq:CGCdRho}) by an additional
scale factor of $\sim1.5$ to return agreement in peripheral collisions; see note 
\footnote{
The impact of collision energy and detector differences on the experimental fitting procedure can
be further investigated by studying the constant offset term in the experimental fit functions. For
example, given arbitrary normalizations $a$ and $b$ of $\rho_2$ and $\rho_1\rho_1$ respectively, the
soft ridge observable is $\Delta\rho/\sqrt{\rho_{\rm ref}} = (a\rho_2-b\rho_1\rho_1)/b\rho_1\rho_1$. The
correlation strength (\ref{eq:Rdef}) then receives a correction following
$\frac{a}{b}\R+\frac{a-b}{b}$ that is energy and detector dependent. If $a\neq b$ then the
correlation strength $\R$ is scaled and a trivial correlation offset is introduced. Thus, studying
the change in offset with respect to collision energy could reveal potential corrections, but this
parameter has not been reported for all collision energies.}.
In Fig.\ref{fig:AmpCompare} we compare calculated ridge amplitudes from (\ref{eq:CGCdRho}) with
measured data for 200 and 62.4 GeV Au+Au collisions from STAR as well as 2.76 TeV Pb+Pb
collisions from ALICE.  In this paper we only compare to measurements from which elliptic flow has been subtracted, because 
we find that fluctuations dominate $v_3$ at STAR.

%
%
\begin{figure}
\includegraphics[scale=0.45]{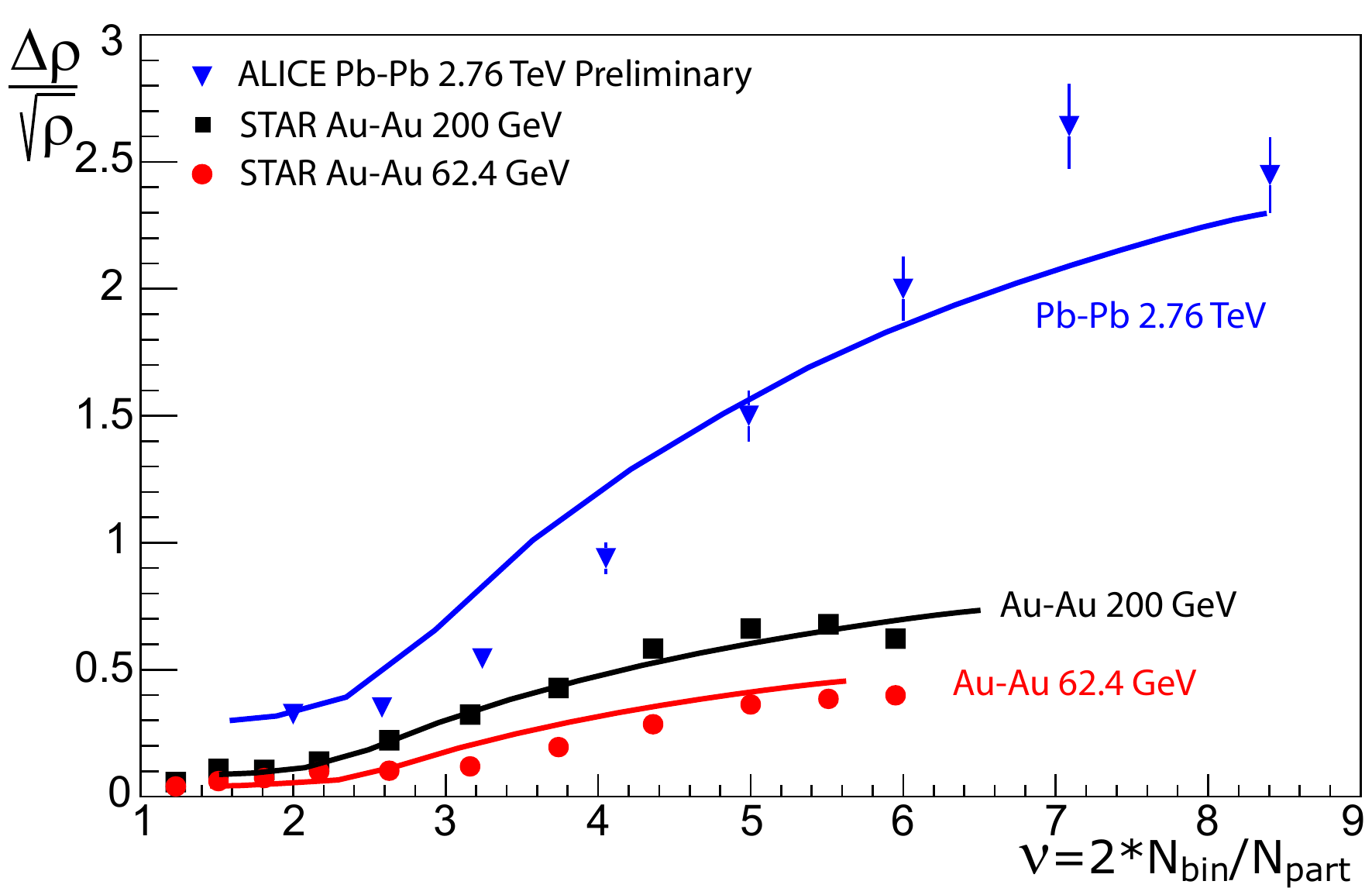}%
\caption{\label{fig:AmpCompare}Near side ridge amplitude calculation from Glasma source correlations. Experimental data is from (STAR)
\cite{Daugherity:2006hz,Daugherity:2008zz,Daugherity:2008su} and (ALICE)
\cite{collaboration:2011um}}
\end{figure}

The characterization of the ridge in terms of Fourier coefficients can prove a valuable tool for
analysis of such discrepancies as well as relate the magnitude its effect with respect to other
phenomena.
Viewing (\ref{eq:sigvn}) in terms of (\ref{eq:sigmaDef}) and (\ref{eq:CGCdRho}), one can write
%
%
\be
2\frac{dN}{dy}\sigma^2_{n} \approx
\int \frac{\Delta\rho(\Delta\phi)}{\sqrt{\rho_{\rm ref}}} \cos (n\Delta\phi)~ d\Delta\phi.
\label{eq:FourierRidge}
\ee
A similar equation was pointed out by Sorensen et. al. \cite{Sorensen:2010zq,Sorensen:2011hm}, and
discussed in terms of event plane eccentricity fluctuations.

We stress, that (\ref{eq:FourierRidge}) is a general result that is model independent. Direct
measurement of this quantity can isolate the effects of flow fluctuations on the ridge by comparing
the results to independent flow fluctuation measurements. Furthermore, flow fluctuation measurements
can be an effective way to quantify the ridge. Deviations from
(\ref{eq:FourierRidge}) at large $\Delta\eta$. can indicate the degree at which other phenomena
contribute to long range correlation measurements. At smaller $\Delta\eta$ shorter range phenomenon
such as jets or diffusion also come into play. Combined flow fluctuation and ridge studies analyzed
with increasing lower \pt limits as done in \cite{DeSilva:2009yy} can reveal the emergence of jets
and their influence. For example, one could test contributions to $v_n\{2\}$ from correlations
induced by jet quenching as suggested in \cite{Moschelli:2009tg,Shuryak:2007fu}.

%
%
\section{\label{sec:factotum}Factorization}
We now ask whether the azimuthal dependence of the pair distribution is entirely determined by the flow coefficients. This is somewhat of a circular question, because flow coefficients are themselves derived from correlation analyses.  A meaningful answer depends on how the flow coefficients are defined. 
Experimentalists expand the transverse-momentum-dependent pair distribution as
\be
\label{eq:rho2fourier2}
\rho_2 \propto 1+ 2 \sum_n V_{n\Delta}(p_{t1}, p_{t2}) \cos(n\Delta\phi).
\ee
The term ``factorization'' is often applied when one can express the Fourier coefficients $V_{n\Delta}(p_{t1}, p_{t2})$ as a product of flow coefficients $v_n(p_{ti})$. Authors often cite factorization as a signature of collective flow 
\cite{Adare:2008ae,Aad:2012bu,Chatrchyan:2012wg,Aamodt:2011by}; see also \cite{Luzum:2011mm} for a theoretical perspective.   
Problems stem from the fact that $V_{n\Delta} \equiv v_n\{2\}^2$ as an exact consequence of the definition (\ref{eq:Dynamic0}). 
In other words, $V_{n\Delta}$  {\em always}  factorizes as $v_n\{2\}\times v_n\{2\}$, regardless of the source of anisotropy.

To define factorization in a useful way, one must compare $V_{n\Delta}$ to the  coefficients $\langle v_n (p_{ti})\rangle$ defined by the reaction plane as in (\ref{eq:vnRP}). Proxies like $v_n\{4\}$ can also be used. To see why this is true, we define  $r_n$ to be the Fourier coefficient of the two particle  correlation function $r(\Delta\phi)$ in (\ref{eq:corrFunExp0}). We then compare  (\ref{eq:rho2fourier2}) to (\ref{eq:rho2fourier}) to find
\be\label{eq:fact}
V_{n\Delta}(p_{t1},p_{t2}) =\langle v_n (p_{t1})\rangle  \langle v_n (p_{t2})\rangle +\frac{ r_n(p_{t1},p_{t2})}{\rho_{\rm ref}(p_{t1},p_{t2})},
\ee
where,  for simplicity, we keep only the leading order in the number of particles in the relevant $p_t$ ranges. The first term in (\ref{eq:fact}) measures the effect of global geometric anisotropy, while the second term is due to local fluctuations.  
Integrating over momenta and using (\ref{eq:vnsigma}), we indeed find 
\be\label{eq:factotum}
V_{n\Delta} =  \langle v_n \rangle^2 + 2 \sigma_n^2  \equiv v_n\{2\}^2,
\ee
as noted above. However, (\ref{eq:fact}) shows that the Fourier coefficients of the correlation function  $r_n(p_{t1},p_{t2})$ break factorization in terms of $\langle v_n (p_{t2})\rangle$.  

We see that hydrodynamic flow does not generally imply factorization, as follows from (\ref{eq:MomCorr}) and (\ref{eq:fact}). Specifically, (\ref{eq:FourierRidge}) shows that flow fluctuations violate factorization. In addition, jets, HBT, and resonances contribute to $r_n$ along with the long range correlations we have considered here. 

Flow coefficients at the LHC are found to factorize for pairs of high $p_t$ particles, but not when high and low $p_t$ particles are mixed \cite{Aamodt:2011by}.  This may seem puzzling if you view factorization as a hydrodynamic signature.  
To understand this result, observe that the contribution $r_n/\rho_{\rm ref}$ in (\ref{eq:fact}) can be large compared to anisotropy only when a substantial fraction of pairs in the selected momentum range are correlated. RHIC energy measurements show that the overall magnitude of the correlation function drops precipitously relative to mixed-event pairs when a lower $p_t$ cutoff is increased above $\sim 1$~GeV \cite{DeSilva:2009yy}.  Our calculations in  \cite{Moschelli:2009tg}, which include both flow and jet-medium interactions, agree with that trend; see, e.g., fig. 9 of \cite{Moschelli:2009tg}. We therefore do not expect significant violations of factorization from the components $r_n$ due to long range correlations from flow or jets above $p_{ti} >  1-2$~GeV.  An analogous trend at higher beam energy may explain the LHC results. 
%
%
\section{\label{sec:summary}Summary}
In this paper we study the connection between flow fluctuations and initial state correlations.  Flow fluctuations are defined by (\ref{eq:sigvn})  using two and four-particle correlation measurements of the harmonic flow coefficients. Section \ref{sec:vn} provides a general discussion of the calculation of flow observables from multiparticle momentum space distributions. Following the cumulant expansion method discussed in \cite{Borghini:2000sa,Borghini:2001vi}, multiparticle distributions can be expanded into factorized and correlated parts, e.g. (\ref{eq:pairDensity}) and (\ref{eq:rho4}); correlated parts contribute directly to the fluctuation observable $\sigma_n$. 

We study the contribution of local long range correlations to $\sigma_n$ based on parton production at common transverse positions. Section \ref{sec:BW} presents calculations of $\sigma_n$ in our Glasma flux tube model. 
We mention that our model incorporates much of the same physics as used by event-by-event hydro in
\cite{Holopainen:2010gz,Petersen:2010cw,Schenke:2010rr,Werner:2010aa,Qiu:2011iv,Mota:2011zz}. Those
authors obtain the initial state in individual events by sampling a probability distribution
analogous to our $\rho_{_{FT}}$. In all cases, Cooper Fry freeze out is used. We use a blast wave
constrained by data to approximate the velocity distribution and the freeze-out surface. The key
differences are that 1) our correlated regions are point-like in the transverse plane, while theirs
may be larger and 2) they evolve their initial distributions using deterministic hydrodynamics in
each event. We expect that they will eventually achieve a better description of the $v_n$ at each
energy. Our complementary aim is to study the big picture by exploring the energy range with a
variety of variables.

The Glasma formulation provides key collision system, energy, and centrality dependences resulting in reasonable agreement with experimental measurements of $v_n\{2\}$ from 62.4 GeV Au+Au to 2.76 TeV Pb+Pb collisions for $n$=2, 3, and 4. The presence of $\sigma_n$ provides two key results: non-zero values of even harmonics in central (circular) collisions, and the existence of $v_3\{2\}$ without triangular flow.

In Sec.\ref{sec:ridge} we turn to discuss two particle correlations and, in particular, the ridge. We see that flow fluctuations and the ridge are facets of the same phenomenon, related by a Fourier transform (\ref{eq:FourierRidge}). The computed peak ridge amplitude in Fig.\ \ref{fig:AmpCompare} is in reasonable agreement with calculations from 62.4 GeV Au+Au to 2.76 TeV Pb+Pb, consistent with our flow results. 
Further investigation is needed to understand the full correlation landscape. In particular, we did not discuss the $n=1$ harmonic, in which momentum conservation plays an important role.  
Characterization of the ridge in terms of flow fluctuations (i.e., harmonics) combined with earlier fit procedures as in Ref.\ \cite{Daugherity:2008su}
will surely prove a useful tool.

We discuss the question of whether the Fourier coefficients of the correlation function factorize into a product of flow coefficients in Sec.\ref{sec:factotum}. In general, irreducible two-particle correlations from any mechanism violate this factorization \cite{Adare:2008ae,Aad:2012bu,Chatrchyan:2012wg}. In particular, the long range correlations that we compute violate factorization at low $p_t$, as would jets, resonance decays, and HBT effects. We expect any such correlations to become negligible for pairs above $p_t \sim 1-2$~GeV, due to the rapid decrease of the number of correlated pairs relative to the mixed-event background that is observed experimentally \cite{DeSilva:2009yy}.  

Finally, we observe that early-time fluctuations have broad implications beyond the flow fluctuations and azimuthal correlations studied here.  In Ref.\  \cite{Gavin:2011gr} we studied the impact of these correlations on multiplicity and $p_t$ fluctuations. We argued in Sec.\ \ref{sec:loco} that $p_t$ and flow fluctuations are intimately related because they are both driven by local hydrodynamic fluctuations.  While the average flow coefficients are primarily determined by the global event shape, the fluctuations of these coefficients have more in common with other fluctuation observables.  Importantly, the overall magnitude of the contribution from long range correlations to all of these fluctuations -- flow, multiplicity, and $p_t$ -- is set by a single scale factor $\cal R$. In Glasma theory, the dependence of $\cal R$ on energy, centrality, and projectile mass are fixed its variation with the saturation scale $Q_s$. Results here and in Ref.\  \cite{Gavin:2011gr} provide a survey of the ridge, and multiplicity, momentum and flow fluctuations that reveals a common energy and centrality dependence that we attribute to the production mechanism. Glasma calculations are consistent with this dependence.
%
%
\begin{acknowledgments}
We thank M.\  Bleicher, H.\ Caines, C.\ DeSilva, J.\ Jia, A.\ Majumder, P.\ Mota, C.\ Pruneau, P.\ Sorensen, R.\ Snellings, A.\ Timmins and S.\ Voloshin for discussions. S.G. thanks the Institute for Nuclear Theory in Seattle, Washington for hospitality during the completion of this work.  
S.G. thanks the Institute for Nuclear Theory in Seattle, Washington for hospitality during the completion of this work. This work was supported in part by the U.S. NSF grant PHY-0855369 (SG) and The Alliance Program of the Helmholtz Association (HA216/EMMI) (GM)
\end{acknowledgments}

%
%
\appendix
\section{\label{appendA}Four-Particle Flow Coefficients}
In this appendix we calculate corrections to the four-particle flow coefficient measurement
$v_n\{4\}$ from two-particle correlation sources. Trying to keep the notation as general as
possible we follow \cite{Borghini:2000sa} using the cumulant expansion method to write
%
%
\bea
&&\la e^{\imath n(\phi_1 - \phi_2)}\ra =
\frac{\int\rho_2(\mathbf{p}_{1},\mathbf{p}_{2})e^{\imath n(\phi_1 - \phi_2)}
d\mathbf{p}_{1}d\mathbf{p}_{2}}
{\int\rho_2(\mathbf{p}_{1},\mathbf{p}_{2})d\mathbf{p}_{1}d\mathbf{p}_{2}}
\label{eq:vn2expo}
\\ \nonumber
\\ 
&&=
\frac{\la N\ra^2 \la v_n\ra^2}{\la N(N-1)\ra} + 
\frac{\int r(\mathbf{p}_1, \mathbf{p}_2)e^{\imath n\Delta\phi}
d\mathbf{p}_{1}d\mathbf{p}_{2}}
{\la N(N-1)\ra},
\label{eq:vn2expo2}
\eea
where $v_n\{2\}^2 = \la e^{\imath n(\phi_1 - \phi_2)}\ra$. Notice that since $\sin(n\Delta\phi)$ is
odd, the second term in (\ref{eq:vn2expo2}) is equivalent to (\ref{eq:sigmaDef}) when
the pair distribution is symmetric about the normal to the reaction plane. 
Based on the definition (\ref{eq:Rdef}) the factor $\la N\ra^2/\la N(N-1)\ra$ in the
first term of (\ref{eq:vn2expo2}) contributes a correction of $1/(1+\R) \approx 1$ as long as the
multiplicity, $N$, is large. For example, in central 200 GeV Au+Au collision we calculate $\R \approx
0.003$. This number will decrease with increasing collision energy but will increase as collisions
become more peripheral. This behavior is discussed in \cite{Gavin:2011gr}.

Our goal is to calculate the four-particle flow coefficient
%
%
\be
v_n\{4\}^4 = 2 \la e^{\imath n(\phi_1 - \phi_2)}\ra^2 - 
\la e^{\imath n(\phi_1 + \phi_2 - \phi_3 - \phi_4)}\ra,
\label{eq:v4expo}
\ee
where the final term is calculated analogously to (\ref{eq:vn2expo}) but from the four-particle
cumulant expansion (\ref{eq:rho4}). Keeping only contributions from two-particle correlations we
have
%
%
\begin{subequations}
\begin{align}
&\la N(N-1)(N-2)(N-3)\ra\la e^{\imath n(\phi_1 + \phi_2 - \phi_3 - \phi_4)}\ra =
\\
&= \la N\ra^4 \la v_n\ra^4
\\
&+
\la N\ra^2 \la v_n\ra^2 \cdot 2{\it Re}\left\{
\int r(\mathbf{p}_1, \mathbf{p}_2)e^{\imath 2n(\Phi-\psi_{_{RP}})}
d\mathbf{p}_{1}d\mathbf{p}_{2} \right\}
\\
&+
4\la N\ra^2\la v_n\ra^2
\int r(\mathbf{p}_1, \mathbf{p}_2)e^{\imath n\Delta\phi}
d\mathbf{p}_{1}d\mathbf{p}_{2}
\label{eq:4sig}
\\
&+
\left|\int r(\mathbf{p}_1, \mathbf{p}_2)e^{\imath 2n(\Phi-\psi_{_{RP}})}
d\mathbf{p}_{1}d\mathbf{p}_{2}
\right|^2
\\
&+
2\left|\int r(\mathbf{p}_1, \mathbf{p}_2)e^{\imath n\Delta\phi}
d\mathbf{p}_{1}d\mathbf{p}_{2} \right|^2.
\label{eq:2sigSq}
\end{align}
\end{subequations}
where $\Delta\phi = \phi_1 - \phi_2$ and $\Phi = (\phi_1 + \phi_2)/2$ are the relative and average
coordinates. To reduce the equations further it is necessary to make the approximation that $\la
N(N-1)(N-2)(N-3)\ra \approx \la N(N-1)\ra^2$ then the terms (\ref{eq:4sig}) and (\ref{eq:2sigSq})
will cancel with corresponding terms emerging from the twice the square of (\ref{eq:vn2expo2}),
leaving the corrections
%
%
\be
v_n\{4\}^4 =
\la v_n\ra^4 -\la v_n\ra^2 \cdot 2{\it Re}\left\{ \Sigma_n^2 \right\} - \left|\Sigma_n^2\right|^2,
\label{eq:vn4correct}
\ee
where
%
%
\be
\Sigma_n^2 = 
\frac{\int r(\mathbf{p}_1, \mathbf{p}_2)
e^{\imath 2n(\Phi-\psi_{_{RP}})}
d\mathbf{p}_{1}d\mathbf{p}_{2}}
{\la N(N-1)\ra}
\ee
which is equivalent to (\ref{eq:BigSigDef}). Notice that these corrections are effectively of order
$2n$ and depend on the reaction plane. As discussed in the text, the corrections in
(\ref{eq:vn4correct}) maximally modify (\ref{eq:v4}) by 1.2\% in our model.

\bibliography{ridge_fluctuations_refs}

\end{document}